\documentclass[reprint,amsmath,amssymb,aps,floatfix]{revtex4-2}
\usepackage{graphicx}
\usepackage{dcolumn}
\usepackage{bm}
\usepackage{hyperref}
\usepackage{lipsum}
\usepackage{subcaption}  
\usepackage{caption}         
\usepackage{float}
\usepackage{tikz}
\usepackage{ragged2e}
\usepackage{multirow}
\usepackage{placeins}
\usetikzlibrary{positioning, arrows.meta}  
\usepackage{relsize}  
\usepackage{overpic}  
\usepackage{placeins}
\usepackage{amssymb}
\usepackage{amsmath}
\usepackage{makecell}
\usepackage{bm}
\usepackage{placeins}
\usepackage[export]{adjustbox}
\usepackage{amsthm}

\theoremstyle{remark}

\usepackage{mathtools}
\usepackage{float}
\usepackage[linesnumbered,ruled,vlined]{algorithm2e}
\usetikzlibrary{positioning, arrows.meta}  
\usepackage{relsize}  
\usepackage{overpic}

\usepackage{tabularx}
\usepackage{booktabs}
\usepackage{array}
\usepackage[ruled,vlined]{algorithm2e}

\usepackage{xcolor}
\definecolor{editgreen}{rgb}{0.00,0.42,0.18}

\begin{document}

\preprint{APS/123-QED}

\title{Label-Permutation Symmetry and Stability in Oscillator Potts Machines}

\author{E.M.~Hasantha~Ekanayake and
Nikhil~Shukla\textsuperscript{\#}}

\affiliation{%
\textnormal{\textsuperscript{}}University of Virginia, Charlottesville, VA, USA\\
\textnormal{\textsuperscript{\#}Email: ns6pf@virginia.edu}
}

\begin{abstract}

Oscillator Potts machines (OPMs) provide a physics-inspired, energy-minimization framework for solving combinatorial optimization problems described by the $q$-state Potts Hamiltonian. Although OPMs may be viewed as multistate extensions of oscillator Ising machines (OIMs), here, we show that they exhibit dynamical properties absent in the binary case. Specifically, we derive a configuration-dependent local-stability condition for a recently proposed multiharmonic OPM formulation and show that configurations with the same Potts energy need not be dynamically equivalent. In particular, for $q\geq4$, permutations of the Potts labels can alter the Jacobian spectrum and, consequently, the regularization strength required to locally stabilize a given Potts configuration. Thus, different phase encodings of the same Potts solution can exhibit different local stability properties despite having identical Potts energies.

\end{abstract}
                         
\maketitle

\section{Introduction}

The increasing memory, computational, and data-movement bottlenecks of conventional digital computing have intensified the search for alternative computing paradigms. In the context of combinatorial optimization, the idea of leveraging the physics of dynamical systems to solve such problems has gained significant attention. The underlying principle behind this computing paradigm is to encode the optimization problem into the dynamics of a physical system. Specifically, a continuous relaxation of the objective function is mapped onto an effective energy function governing the system dynamics. Consequently, as the system relaxes toward lower-energy states, its dynamics naturally ``compute'' candidate low-energy solutions to the underlying optimization problem \cite{Wang2021, ekanayake2026mindgapanalogising,Kalinin2022,ErcseyRavasz2011, Potts_model}.

Oscillator Ising machines (OIMs) ~\cite{Wang2021,TodriSanial2024} represent a classic realization of this principle. In an OIM, binary variables are encoded in two phase states (typically, $\theta \in \{0,\pi\}$) of a network of coupled oscillators, and the oscillator interactions are designed such that the corresponding phase configurations represent low-energy states of an Ising Hamiltonian. OIMs have been investigated in both optical and electronic platforms using a variety of oscillator and coupling mechanisms \cite{Mohseni2022}. Since the Ising spins satisfy \(\sigma_i\in\{-1,+1\}\) (equivalent binary encoding $s_i=0.5(1+\sigma_i)$), these systems can direclty represent optimization problems with two discrete states per spin (e.g., MaxCut).

A natural generalization is obtained by allowing each variable to occupy one of \(q\geq 2\) discrete states, leading to the \(q\)-state Potts model \cite{RevModPhys.54.235}. The Ising model is then recovered as the special case \(q=2\). Similar to OIMs, Oscillator Potts machines (OPMs) aim to provide a physical realization of the $q$-state Potts model by encoding the Potts configurations in \(q\) oscillator phase states \cite{HonariLatifpourMiri+2020+4199+4205}, which subsequently, enables direct mapping of combinatorial optimization such as graph coloring and clustering which may require the graph node to be assigned $q\geq 2$ states. This generalization, however, introduces unique dynamical features that are absent in the binary case because the local stability of a phase configuration can depend not only on whether neighboring variables occupy the same state, but also on the relative arrangement of the assigned states (``labels'') on the phase circle.

In this work, we develop a linear-stability framework for OPMs governed by the dynamics proposed by Cheng and Lin~\cite{cheng2026isingpottsphysicsinspiredpotts}. We derive the stability threshold for an arbitrary discrete \(q\)-state configuration and examine how it changes under permutations of the Potts labels. For \(q=2\) (Ising) and \(q=3\), every label permutation leaves the interaction Jacobian unchanged, preserving its spectrum and stabilization threshold. For \(q\geq4\), a permutation can change the Jacobian, and in some cases, its largest eigenvalue. Consequently, two labelings of the same Potts configuration can have identical Potts energy but different stabilization thresholds.

\section{Stability Thresholds in OPMs}

We begin by considering an OPM consisting of \(N\) coupled oscillators, whose phases evolve according to the dynamics introduced by Cheng and Lin~\cite{cheng2026isingpottsphysicsinspiredpotts}:
\begin{equation}
\dot{\theta}_i
=
-K  \sum_{j=1}^{N} J_{ij}
\sum_{m=1}^{q-1}
m(q-m)
\sin\!\left[m(\theta_i-\theta_j)\right]
-
K_{\mathrm{s}}\sin(q\theta_i).
\label{eq:opm_deterministic}
\end{equation}
Here, \(\theta_i\in[0,2\pi)\) denotes the phase of oscillator \(i\), \(J_{ij}\) denotes the interaction weight between oscillators \(i\) and \(j\), \(K\) sets the overall interaction strength, and \(K_{\mathrm{s}}\) controls the strength of the \(q^{\mathrm{th}}\)-harmonic regularization. The interaction function contains harmonics up to order \(q-1\), while the regularization term creates uniformly spaced discrete phase states,
\begin{equation}
\theta_i^{(s_i)}
=
\frac{2\pi s_i}{q},
\qquad
s_i\in\{0,1,\ldots,q-1\}.
\label{eq:potts_phase_states}
\end{equation}

\begin{figure*}[t]
    \centering
    \includegraphics[width=1\linewidth]{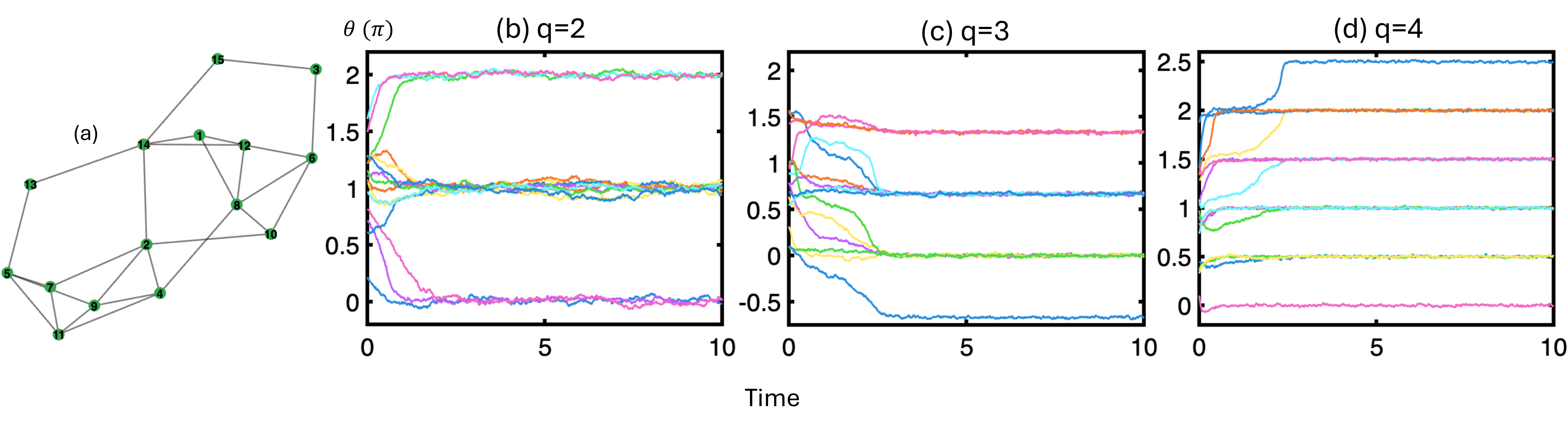}
\caption{\justifying
Stochastic OPM dynamics.
(a) The randomly generated 15-node graph with antiferromagnetic
interactions considered in the simulations.
The corresponding stochastic phase dynamics are shown for
(b) $q=2$, (c) $q=3$, and (d) $q=4$.
The phases are shown in units of $\pi$, such that the encoded states are
$\{0,1\}$ for $q=2$, $\{0,2/3,4/3\}$ for $q=3$, and
$\{0,1/2,1,3/2\}$ for $q=4$.
For each case, the dynamics converge to a ground-state configuration of
the corresponding Potts problem, with ground-state energies
$E_{\mathrm{gs}}=6$, $0$, and $0$ for $q=2$, $3$, and $4$,
respectively. The ground-state energies were independently verified
using MATLAB\textsuperscript{\textregistered}'s \texttt{intlinprog} mixed-integer linear programming
(MILP) solver \cite{MATLAB2026}.
}
\label{app:fig:simulations}
\end{figure*}

These discrete phase states establish the correspondence between the oscillator phases and the \(q\)-state Potts variables, with each phase state representing one of the \(q\) possible Potts labels. While Eq.~\eqref{eq:opm_deterministic} describes the deterministic OPM dynamics, our numerical simulations additionally incorporate additive noise to
facilitate exploration of the energy landscape and reduce trapping in
local optima, with the corresponding stochastic OPM dynamics given in
Appendix~\ref{app:stochastic_opm}. 

Fig.~\ref{app:fig:simulations} shows numerical simulations of the
stochastic OPM dynamics for $q=2$, $q=3$, and $q=4$, illustrating
the dynamical behavior of the OPM for different numbers of discrete
Potts phase states. Specifically, we consider a randomly generated
15-node graph with antiferromagnetic interactions (Fig.~\ref{app:fig:simulations}(a)). For each value of $q$, the oscillator
phases (unwrapped) evolve toward the corresponding $q$ discrete phase states and
converge to a ground-state configuration of the associated Potts
problem. The resulting ground-state energies are
$E_{\mathrm{gs}}=6$, $0$, and $0$ for $q=2$, $3$, and $4$,
respectively, and were independently verified using MATLAB\textsuperscript{\textregistered}'s
\texttt{intlinprog} mixed-integer linear programming (MILP)
solver~\cite{MATLAB2026}.

To analyze the local stability of the discrete phase configurations, we
assume symmetric interactions, \(J_{ij}=J_{ji}\). At a discrete Potts
configuration, Eq.~\eqref{eq:potts_phase_states} gives
\begin{equation}
\theta_i^{*}-\theta_j^{*}
=
\frac{2\pi}{q}(s_i-s_j).
\end{equation}
We therefore introduce the relative label difference
\begin{equation}
r_{ij}
=
s_i-s_j \pmod q,
\qquad
r\in\mathbb{Z}_q,
\label{eq:relative_label_difference}
\end{equation}
and define the corresponding interaction coefficient, which is equivalent to the local curvature of the interaction energy.

\begin{equation}
\kappa_q(r_{ij})
=
\sum_{m=1}^{q-1}
m^2(q-m)
\cos\!\left(m\frac{2\pi r_{ij}}{q}\right),
\qquad
r_{ij}\in\mathbb{Z}_q.
\label{eq:potts_kappa_gap}
\end{equation}
At the discrete Potts configurations, \(\cos(q\theta_i^{*})=1\). Therefore the Jacobian of
Eq.~\eqref{eq:opm_deterministic}, evaluated at a discrete 
configuration \(\bm{\theta}^{*}\), has entries
\begin{equation}
\mathcal{J}_{q{\{i,j\}}}(\bm{\theta}^{*})
=
\begin{cases}
\displaystyle
-K\sum_{k=1}^{N}J_{ik}\kappa_q(r_{ik})
-qK_{\mathrm{s}},
& i=j,\\[3mm]
\displaystyle
KJ_{ij}\kappa_q(r_{ij}),
& i\neq j,
\end{cases}
\label{eq:opm_jacobian_discrete}
\end{equation}
where the terms proportional to \(K\) arise from the oscillator
interactions, and the term \(-qK_{\mathrm{s}}\) on the
diagonal arises from the \(q^{\mathrm{th}}\)-harmonic
regularization. The interaction part of the Jacobian has a natural elementary Laplacian
structure \cite{tropp2019matrix}. For each interacting edge \(\{i,j\}\in E\), define the
elementary Laplacian 
\begin{equation}
L_{ij}
=
(\bm e_i-\bm e_j)
(\bm e_i-\bm e_j)^{\mathsf T},
\label{eq:potts_edge_laplacian_gap}
\end{equation}
where \(\bm e_i\) denotes the \(i^{\mathrm{th}}\) standard basis vector
in \(\mathbb{R}^{N}\). The Jacobian can then be expressed compactly as
\begin{equation}
\mathcal{J}_q(\bm{\theta}^{*})
=
-K\sum_{\{i,j\}\in E}
J_{ij}\kappa_q(r_{ij})L_{ij}
-qK_{\mathrm{s}}I.
\label{eq:opm_jacobian_edge}
\end{equation}
For a discrete Potts configuration
\(\bm{s}=(s_1,\ldots,s_N)\), defining \textit{ interaction Jacobian}
\begin{equation}
B_q(\bm{s})
=
-\sum_{\{i,j\}\in E}
J_{ij}\kappa_q(s_i-s_j)L_{ij}.
\label{eq:interaction_jacobian_uniform_sign}
\end{equation}
The full Jacobian
can then be written as
\begin{equation}
\mathcal{J}_q(\bm{s})
=
K B_q(\bm{s})
-
qK_s I.
\label{eq:full_potts_jacobian}
\end{equation}

Therefore, local asymptotic stability of a given Potts configuration requires all eigenvalues of
\(\mathcal{J}_{q}(\bm s)\) to be negative, yielding the stabilization
threshold
\begin{equation}
    K_s^{\mathrm{stab}}(\bm s)
    =
    \frac{K}{q}
    \lambda_{\max}\!\left[
        B_{q}(\bm s)
    \right].
    \label{eq:general_potts_threshold_revised}
\end{equation}
Thus, the local stability of a discrete Potts configuration is determined by the interaction coefficients \(\kappa_q(s_i-s_j)\), which depend on the difference between the Potts labels at
the two ends of each edge, together with the graph structure encoded by the elementary Laplacians \(L_{ij}\).
\\

\paragraph*{Potts ground-state stabilization threshold---}
For any discrete Potts configuration $\bm s$, each edge Laplacian satisfies 
\begin{equation}
    L_{ij}\bm 1 = 0.
\end{equation}
It follows directly from Eq.~\eqref{eq:interaction_jacobian_uniform_sign}
that
\begin{equation}
    B_{q}(\bm s)\bm 1 = 0.
\end{equation}
Thus, \(\bm{1}\) is an eigenvector of \(B_q(\bm{s})\) associated with the zero eigenvalue, meaning that
\begin{equation}
    0\in\operatorname{spec}
    \left[B_{q}(\bm s)\right].
\end{equation}
Since $B_{q}(\bm s)$ is symmetric, all of its eigenvalues
are real, and therefore
\begin{equation}
    \lambda_{\max}
    \left[B_{q}(\bm s)\right]
    \geq 0.
\end{equation}

\noindent Using Eq.~\eqref{eq:general_potts_threshold_revised}, the stabilization
threshold of any discrete Potts configuration consequently satisfies
\begin{equation}
    K_s^{\mathrm{stab}}(\bm s)
    =
    \frac{K}{q}
    \lambda_{\max}
    \left[B_{q}(\bm s)\right]
    \geq 0.
\end{equation}

\noindent Let
\begin{equation}
    \mathcal G_q
    =
    \underset{\bm s\in\mathbb Z_q^N}
    {\operatorname{arg\,min}}\,
    H_{\mathrm{Potts}}(\bm s)
\end{equation}
denote the set of Potts \emph{ground-state} configurations. The onset of
ground-state stabilization is then
\begin{equation}
    K_{s,\mathrm{gs}}^{(q)}
    =
    \min_{\bm s\in\mathcal G_q}
    K_s^{\mathrm{stab}}(\bm s)
    =
    \frac{K}{q}
    \min_{\bm s\in\mathcal G_q}
    \lambda_{\max}
    \left[B_{q}(\bm s)\right].
\end{equation}
Hence,
\begin{equation}
    K_{s,\mathrm{gs}}^{(q)}\geq0.
\end{equation}

\begin{figure}
    \centering
    \includegraphics[width=1\linewidth]{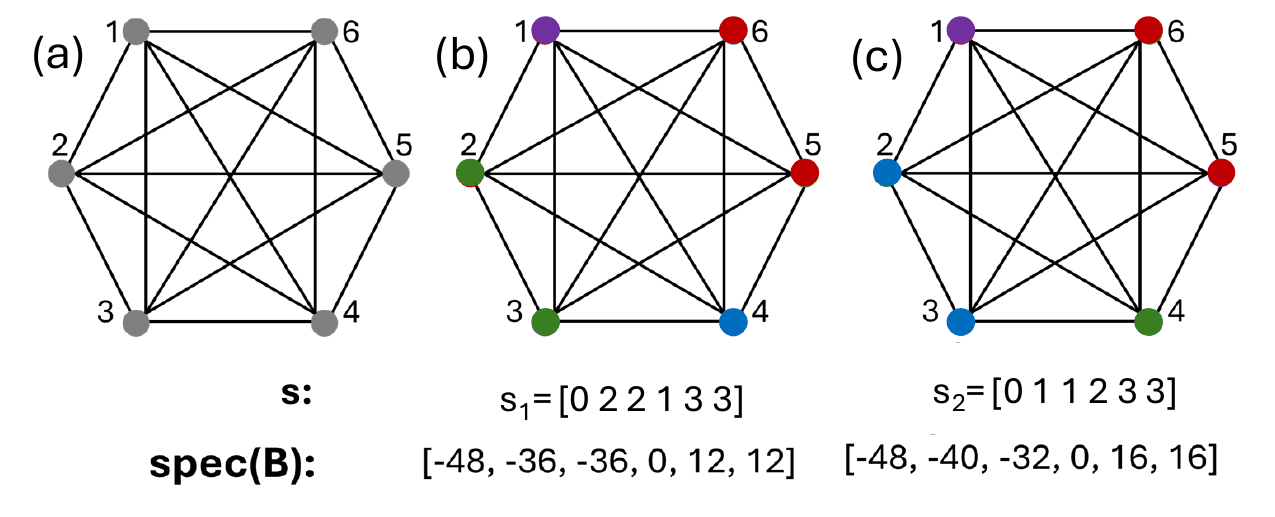}
    \caption{\justifying
    Dynamically in-equivalent phase encodings of the same Potts ground state partition. (a) 6 node complete graph with negative (anti-ferromagnetic) coupling. The middle (b); and right (c) panels represent the same $q=4$ ground-state Potts    partition. The only modification is the interchange of the blue and green Potts labels, while the red and purple labels remain unchanged. Although this relabeling leaves the Potts energy unchanged, it changes the edgewise phase separations and hence the normalized interaction Jacobian. The resulting Jacobian spectra are different, with largest eigenvalues $12$ and $16$.}
    \label{fig:phase_label}
\end{figure}

\section{Potts Label Permutations}
\subsection*{Dynamically inequivalent representations of the same Potts ground state}

An important consequence of the OPM phase encoding is that the dynamical stability of a Potts solution depends not only on the underlying graph partition, but also on the assignment of the Potts labels to the discrete oscillator phases. The discrete Potts Hamiltonian, $H_{\mathrm{Potts}}(\bm{s})=-\sum_{i<j}^{N}J_{ij}\delta_{s_i,s_j}$,
where $\delta_{s_i,s_j}$ is the Kronecker delta, is invariant under permutations of the \emph{q} Potts labels. Therefore, different labelings of the same partition represent the same combinatorial solution and have identical energy. However, from Eq.~\eqref{eq:interaction_jacobian_uniform_sign}, the OPM interaction Jacobian depends explicitly on the edgewise label differences through $\kappa_q(s_i-s_j)$. Consequently, a permutation of the phase labels can modify the spectrum of $B_{q}(\bm s)$, and hence the stabilization threshold, without changing the corresponding Potts solution.

We illustrate this with an example. In Fig.~\ref{fig:phase_label} we consider antiferromagnetic coupling and show the dependence of local stability on the assignment of Potts labels for $q=4$. The \(q=4\) states are represented by colors, violet, blue, green and red, which represent the phases $\theta=0,\pi/2,\pi, \text{and } 3\pi/2$, respectively. The two configurations shown in Fig.~\ref{fig:phase_label}(b) and (c) have the same ground-state partition: $\{1\}\{2,3\}\{4\}\{5,6\}$; the only modification is that the blue and green Potts labels assigned to the corresponding groups are interchanged, while the red and purple labels remain unchanged. Consequently, the two configurations have identical Potts energy. Their local stability, however, is different because the interaction Jacobian depends explicitly on the relative phase-label differences across the edges through $\kappa_q(s_i-s_j)$. Interchanging the blue and green labels therefore changes the edgewise phase separations and, consequently, the Jacobian spectrum. 

\begin{figure}[t]
    \centering
    \includegraphics[width=0.8\columnwidth]{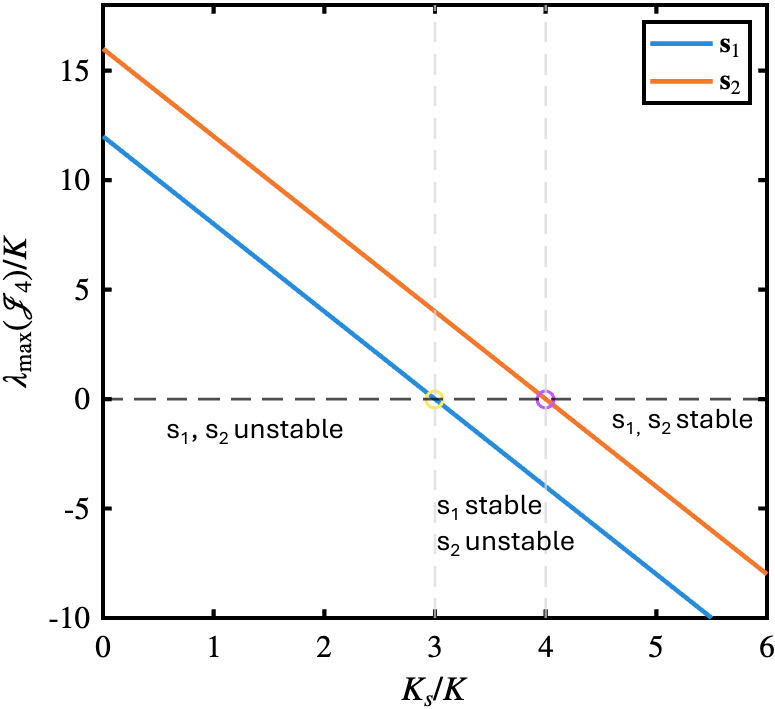}
    \caption{ \justifying
    Configuration-dependent stabilization thresholds for the two permutation-equivalent $q=4$ Potts ground states shown in Fig.~\ref{fig:phase_label}. The dominant eigenvalues of the full OPM Jacobian, $\lambda_{\max}(\mathcal{J}_4)/K$, are shown as functions of the normalized regularization strength $K_s/K$. The zero crossings at $K_s/K=3$ and $K_s/K=4$ correspond to the stabilization thresholds of $\bm{s}_1$ and $\bm{s}_2$, respectively. Positive values indicate local instability, whereas negative values indicate local asymptotic stability.}
    \label{fig:stability_verification}
\end{figure}

For the two assignments shown in Figs.~\ref{fig:phase_label}(b) and
\ref{fig:phase_label}(c), the largest eigenvalues of the interaction
Jacobians are
\begin{equation}
    \lambda_{\max}
    \left[B_{4}(\bm s_1)\right]
    =12,
    \qquad
    \lambda_{\max}
    \left[B_{4}(\bm s_2)\right]
    =16,
\end{equation}
respectively. Using Eq.~\eqref{eq:general_potts_threshold_revised}, the
corresponding stabilization thresholds are
\begin{equation}
    K_s^{\mathrm{stab}}(\bm s_1)
    =
    \frac{K}{4}(12)
    =
    3K,
    \qquad
    K_s^{\mathrm{stab}}(\bm s_2)
    =
    \frac{K}{4}(16)
    =
    4K.
    \label{eq:thresholds_Ks}
\end{equation}
Thus, even within the same Potts ground-state partition, different
assignments of Potts labels to oscillator phases can lead to distinct
Jacobian spectra and, consequently, different stabilization thresholds.
Therefore, combinatorially equivalent ground states are not necessarily
dynamically equivalent under the OPM phase encoding.
\\

\textit{Configuration-dependent stability---}
The difference between the stabilization thresholds of $\bm{s}_1$ and
$\bm{s}_2$ can be further illustrated by examining the dominant
eigenvalues of their full Jacobians as the normalized regularization
strength $K_s/K$ is varied. Using
Eq.~\eqref{eq:full_potts_jacobian}, the dominant Jacobian eigenvalues
are
\begin{equation}
\frac{\lambda_{\max}[\mathcal{J}_4(\bm{s}_1)]}{K}
=12-4\frac{K_s}{K},
\qquad
\frac{\lambda_{\max}[\mathcal{J}_4(\bm{s}_2)]}{K}
=16-4\frac{K_s}{K}.
\end{equation}
As shown in Fig.~\ref{fig:stability_verification}, these eigenvalues cross zero at $K_s/K=3$ ($K_s=3K$) and
$K_s/K=4$ ($K_s=4K$), respectively, consistent with
the stabilization thresholds derived in Eq. \ref{eq:thresholds_Ks}. These zero crossings
define three distinct stability regimes. For $K_s/K<3$, both
configurations are unstable. In the intermediate regime
$3<K_s/K<4$, $\bm{s}_1$ is locally stable, whereas $\bm{s}_2$ remains
unstable. For $K_s/K>4$, both configurations are locally stable.
Thus, the two label-permutation-equivalent Potts ground states exhibit
different stability regimes as the regularization strength is varied.

\subsection*{Theoretical Framework} 

\noindent We now establish the underlying theoretical framework for how different label permutations can lead to different stability properties. Let
\begin{equation}
\bm{s}=(s_1,\ldots,s_N),
\qquad
s_i\in\{0,\ldots,q-1\},
\end{equation}
denote a $q$-state Potts configuration. Let
$P:S\rightarrow S$ denote an arbitrary permutation (relabeling) of the Potts
state set
\[
S=\{0,\ldots,q-1\}.
\]
The relabeled configuration is obtained by applying $P$ independently to the
state of each node,
\begin{equation}
(P\bm{s})_i=P(s_i),
\qquad i=1,\ldots,N.
\label{eq:relabel}
\end{equation}

\noindent \textit{Permutation Symmetry of the Potts Hamiltonian---} The discrete Potts Hamiltonian depends only on whether the states assigned
to two neighboring nodes are equal. It can be written as
\begin{equation}
 H_{\mathrm{Potts}}(\bm s)
    =
    -\sum_{i<j}^NJ_{ij}\delta_{s_i,s_j},
\label{eq:potts_hamiltonian}
\end{equation}
where $\delta_{s_i,s_j}$ denotes the Kronecker delta. Since $P$ is a bijection on the Potts state set,\,
\begin{equation}
\delta_{P(s_i),P(s_j)}
=
\delta_{s_i,s_j}.
\end{equation}
The Potts Hamiltonian is invariant under arbitrary permutations of the
Potts-state labels~\cite{Arovas2019}. Accordingly:
\begin{equation}
H_{\mathrm {Potts}}(P\bm{s})
=
H_{\mathrm {Potts}}(\bm{s}),
\qquad
\forall\,P\in S_q.
\label{eq:potts_permutation_invariance}
\end{equation}
Equation~\eqref{eq:potts_permutation_invariance} implies that the discrete
Potts Hamiltonian possesses the full color-permutation symmetry $S_q$.
Consequently, a relabeling of the Potts states does not change the underlying
graph partition or the value of the Potts Hamiltonian.\\

\noindent \textit{Permutation Dependence of the OPM Jacobian---}
Unlike the discrete Potts Hamiltonian, which depends only on whether two
neighboring nodes occupy the same state, the OPM interaction Jacobian
depends on the complete discrete phase separation between neighboring
Potts states. As derived in Eq.~\eqref{eq:interaction_jacobian_uniform_sign}, the
interaction Jacobian can be expressed in the elementary Laplacian form,

\begin{equation}
    B_{q}(\bm s)
    =
    -
    \sum_{\{i,j\}\in E}
    J_{ij}\kappa_q(s_i-s_j)L_{ij}.
    \label{eq:appendix -interaction_jacobian_uniform_sign}
\end{equation}
After applying a permutation $P$ to the Potts labels, the interaction
Jacobian becomes
\begin{equation}
B_q(P\bm{s})
=
-
\sum_{\{i,j\}\in E}
J_{ij}\,
\kappa_q\!\left(
P(s_i)-P(s_j)
\right)
L_{ij}.
\label{eq:opm_jacobian_permuted}
\end{equation}
Here the interaction coefficient
$\kappa_q(r_{ij})$ depends on the complete discrete phase separation between
neighboring Potts states. A relabeling of the Potts states can modify the interaction
coefficient associated with one or more graph edges. Specifically,
if
\begin{equation}
\Delta\kappa_{ij}^{(P)}
=
\kappa_q\!\left(P(s_i)-P(s_j)\right)
-
\kappa_q(s_i-s_j)
\neq 0
\label{eq:delta_kappa_permutation}
\end{equation}
for at least one edge \(\{i,j\}\in E\), the relabeling changes the
interaction contribution associated with that edge and can therefore
produce a different Jacobian matrix.
\begin{equation}
B_{q}(P\bm{s})
\neq
B_{q}(\bm{s}).
\label{eq:jacobian_not_invariant}
\end{equation}

\noindent However, a change in the interaction Jacobian does not necessarily
imply a change in its eigenvalue spectrum.
In particular, two distinct interaction Jacobians may have the same
largest eigenvalue,
\begin{equation}
\lambda_{\max}\!\left[B_q(P\bm{s})\right]
=
\lambda_{\max}\!\left[B_q(\bm{s})\right],
\end{equation}
and therefore the same stabilization threshold. Thus, a change in the
interaction coefficients under relabeling is not, by itself, sufficient
to establish dynamically distinct stability thresholds.

Accordingly, permutation-equivalent Potts configurations can fall into two
dynamical classes: some relabelings modify the Jacobian while preserving
its spectrum, whereas others modify both the Jacobian and its spectrum.
Only the latter lead to different OPM stabilization thresholds.\\

\noindent \textit{Jacobian Change Induced by a Potts-Label Permutation---} To quantify the effect of a Potts-label permutation on the OPM dynamics,
we consider the difference between the corresponding interaction Jacobians,
\begin{align}
\Delta B_{q}
&=
B_q(P\bm{s})-B_q(\bm{s})
\nonumber\\
&=
-
\sum_{\{i,j\}\in E}
J_{ij}
\left[
\kappa_q\!\left(
P(s_i)-P(s_j)
\right)
-
\kappa_q(s_i-s_j)
\right]
L_{ij}
\nonumber\\
&=
-
\sum_{\{i,j\}\in E}
J_{ij}\,
\Delta\kappa_{ij}^{(P)}
L_{ij},
\label{eq:delta_B_compact}
\end{align}
where $\Delta\kappa_{ij}^{(P)}
=
\kappa_q\!\left(P(s_i)-P(s_j)\right)
-
\kappa_q(s_i-s_j)$. 
Equation~\eqref{eq:delta_B_compact} shows that only those edges whose
interaction coefficient changes under the relabeling contribute to the
Jacobian perturbation.

\subsubsection*{Sufficient Condition for Jacobian Invariance}

A sufficient condition for the interaction Jacobian to remain unchanged is
\begin{equation}
\Delta\kappa_{ij}^{(P)}
=
0,
\qquad
\forall\,\{i,j\}\in E,
\label{eq:jacobian_invariant_condition}
\end{equation}
which implies
\begin{equation}
\Delta B_{q}=0.
\end{equation}

\subsubsection*{Conditions for a Change in the Stabilization Threshold}

A necessary condition for a Potts-label permutation to modify the
interaction Jacobian is that there exists at least one graph edge for
which the interaction coefficient changes, i.e.,
$\Delta\kappa_{ij}^{(P)}\neq0$.

When this condition is satisfied, the interaction Jacobian may change.
However, a change in the Jacobian does not necessarily imply a change in
its eigenvalue spectrum,  since distinct Jacobian matrices can be isospectral. In particular, because the
stabilization threshold is determined by the largest eigenvalue, a change
in the Jacobian alone is insufficient to establish a change in dynamical
stability.

To determine when the stabilization eigenvalue changes, consider two
Potts configurations related by a label permutation, with
\begin{equation}
    B_1=B_{q}(\bm{s}),
    \qquad
    B_2=B_{q}(P\bm{s}),
    \qquad
    \Delta B_q=B_2-B_1.
\end{equation}
Let $\bm v_1$ and $\bm v_2$ denote normalized dominant eigenvectors of
$B_1$ and $B_2$, respectively,
\begin{align}
    B_1\bm v_1
    &=
    \lambda_{\max}(B_1)\bm v_1,
    &
    \|\bm v_1\|&=1,
    \\
    B_2\bm v_2
    &=
    \lambda_{\max}(B_2)\bm v_2,
    &
    \|\bm v_2\|&=1.
\end{align}
Applying the Rayleigh--Ritz characterization to $B_2$ using $\bm v_1$
gives
\begin{align}
    \lambda_{\max}(B_2)
    &\geq
    \bm v_1^{T}B_2\bm v_1
    \nonumber\\
    &=
    \lambda_{\max}(B_1)
    +
    \bm v_1^{T}\Delta B_q\bm v_1.
\end{align}
Therefore, if
\begin{equation}
    \bm v_1^{T}\Delta B_q\bm v_1>0
    \quad\Longrightarrow\quad
    \lambda_{\max}(B_2)>
    \lambda_{\max}(B_1).
    \label{eq:sufficient_increase}
\end{equation}
Similarly, applying Rayleigh--Ritz to $B_1$ using $\bm v_2$ yields
\begin{align}
    \lambda_{\max}(B_1)
    &\geq
    \bm v_2^{T}B_1\bm v_2
    \nonumber\\
    &=
    \lambda_{\max}(B_2)
    -
    \bm v_2^{T}\Delta B_q\bm v_2,
\end{align}
so that
\begin{equation}
    \bm v_2^{T}\Delta B_q\bm v_2<0
    \quad\Longrightarrow\quad
    \lambda_{\max}(B_2)<
    \lambda_{\max}(B_1).
    \label{eq:sufficient_decrease}
\end{equation}
Thus, a change in the edgewise interaction coefficient is necessary for
the Jacobian to change, whereas the Rayleigh quotients of
$\Delta B_q$ along the dominant eigenvectors provide sufficient
conditions for the largest eigenvalue, and consequently the OPM
stabilization threshold, to increase or decrease under a Potts-label
permutation.\\

\noindent \textit{Illustration using the $q=4$ example---}
We illustrate the above condition using the two permutation-equivalent
$q=4$ Potts ground states shown in Figs.~\ref{fig:phase_label}(b) and (c). For these configurations,
\begin{equation}
\lambda_{\max}(B_1)=12,
\qquad
\lambda_{\max}(B_2)=16.
\end{equation}
Evaluating the Jacobian perturbation
$\Delta B_q=B_2-B_1$ along a normalized dominant eigenvector
$\bm v_1$ of $B_1$ gives
\begin{equation}
\bm v_1^{T}\Delta B_q\bm v_1=4>0.
\end{equation}
Therefore, Eq.~\eqref{eq:sufficient_increase} yields
\begin{equation}
\lambda_{\max}(B_2)
\geq
\lambda_{\max}(B_1)+4
=
16.
\end{equation}
Since direct evaluation gives $\lambda_{\max}(B_2)=16$, the bound is
attained exactly. Consequently, according to Eq. \ref{eq:thresholds_Ks} the stabilization threshold increases
from
\begin{equation}
K_s^{\mathrm{stab}}(\bm{s}_1)=3K
\qquad\text{to}\qquad
K_s^{\mathrm{stab}}(\bm{s}_2)=4K.
\end{equation}
The corresponding interaction Jacobians and the complete numerical
evaluation of the Rayleigh--Ritz condition are provided in
Appendix~\ref{app:q4_rayleigh_ritz}.
\\

\noindent \textit{Dependence of OPM Jacobian Permutation Symmetry on q---} The interaction coefficient introduced in
Eq.~\ref{eq:potts_kappa_gap} is given by
\begin{equation}
\kappa_q(r_{ij})
=
\sum_{m=1}^{q-1}
m^2(q-m)
\cos\!\left(\frac{2\pi mr_{ij}}{q}\right),
\qquad
r_{ij}\in\mathbb{Z}_q.
\label{eq:potts_kappa_gap_1}
\end{equation}
This expression admits the closed form (detailed derivation is given in Appendix \ref{app:kappa_derivation})
\begin{equation}
\kappa_q(r_{ij})
=
\begin{cases}
\displaystyle
\frac{q^2(q^2-1)}{12},
& r_{ij}=0,\\[2mm]
\displaystyle
-\frac{q^2}{4\sin^2\!\left(\pi r_{ij}/q\right)},
& r_{ij}\neq 0.
\end{cases}
\label{eq:kappa_closed_form}
\end{equation}
In particular, for nonzero discrete phase separations,
\begin{equation}
\kappa_q(r_{ij})
=
-\frac{q^2}
{4\sin^2\!\left(\dfrac{\pi r_{ij}}{q}\right)},
\qquad
r_{ij}=1,\ldots,q-1.
\label{eq:appendix_kappa_closed_form}
\end{equation}

\paragraph{Case $q=2$.} For $q=2$, there is only one nonzero phase-separation class,
\[
r_{ij}=1.
\]
Hence,
\begin{equation}
\kappa_2(1)
=
-\frac{4}
{4\sin^2(\pi/2)}
=
-1.
\end{equation}
Since every unequal pair of Potts states has the same interaction
coefficient, every permutation of the Potts labels preserves the coefficient
associated with every graph edge. Therefore,
\begin{equation}
\kappa_2\!\left(P(s_i)-P(s_j)\right)
=
\kappa_2(s_i-s_j),
\qquad
\forall\,\{i,j\}\in E,
\end{equation}
which immediately implies
\begin{equation}
B_{2}(P\bm{s})
=
B_{2}(\bm{s}).
\end{equation}
Consequently,
\begin{equation}
\operatorname{spec}
\!\left[
B_{2}(P\bm{s})
\right]
=
\operatorname{spec}
\!\left[
B_{2}(\bm{s})
\right].
\end{equation}

\paragraph{Case $q=3$.}

For $q=3$, there are two nonzero phase-separation classes,
\[
r_{ij}=1,
\qquad
r_{ij}=2.
\]
Using Eq.~\eqref{eq:kappa_closed_form},
\begin{align}
\kappa_3(1)
&=
-\frac{9}
{4\sin^2(\pi/3)}
=
-3,
\\
\kappa_3(2)
&=
-\frac{9}
{4\sin^2(2\pi/3)}
=
-3.
\end{align}
Since $
\kappa_3(1)=\kappa_3(2)$,
all unequal Potts-state pairs possess the same interaction coefficient.
Therefore,
\begin{equation}
\kappa_3\!\left(P(s_i)-P(s_j)\right)
=
\kappa_3(s_i-s_j),
\qquad
\forall\,\{i,j\}\in E,
\end{equation}
and consequently,
\begin{equation}
B_{3}(P\bm{s})
=
B_{3}(\bm{s}).
\end{equation}
Hence,
\begin{equation}
\operatorname{spec}
\!\left[
B_{3}(P\bm{s})
\right]
=
\operatorname{spec}
\!\left[
B_{3}(\bm{s})
\right].
\end{equation}

\paragraph{Case $q\ge4$.}

For $q\ge4$, there exist at least two distinct nonzero phase-separation
classes. In particular,
\begin{equation}
\kappa_q(1)
=
-\frac{q^2}
{4\sin^2(\pi/q)},
\qquad
\kappa_q(2)
=
-\frac{q^2}
{4\sin^2(2\pi/q)}.
\end{equation}
Since
\begin{equation}
\sin\!\left(\frac{\pi}{q}\right)
\neq
\sin\!\left(\frac{2\pi}{q}\right),
\qquad
q\ge4,
\end{equation}
it follows that
\begin{equation}
\kappa_q(1)
\neq
\kappa_q(2).
\end{equation}

Therefore, a permutation of the Potts labels can change an edge from one
phase-separation class to another, thereby changing its interaction
coefficient. Consequently, the OPM interaction Jacobian is no longer
invariant under the full permutation group:
\begin{equation}
B_{q}(P\bm{s})
\neq
B_{q}(\bm{s})
\end{equation}
for some permutations $P\in S_q$. This is also evident from the example considered in Fig. 1.

Thus, the OPM interaction Jacobian is invariant under arbitrary Potts-label
permutations only for
\begin{equation}
q=2,\;3.
\end{equation}
For $q\ge4$, permutation-equivalent Potts configurations can have different
interaction Jacobians and, consequently, may exhibit different dynamical
stability properties.

\subsection*{Symmetry Reduction in the Oscillator Potts Representation}

The preceding results can also be understood in terms of a
mismatch between the symmetries of the discrete Potts
Hamiltonian and its continuous oscillator realization.
While the discrete Potts Hamiltonian is invariant under
the full label-permutation group $S_q$ \cite{Arovas2019}, the oscillator
interaction coefficient depends on the angular separation
between the encoded phase states. Consequently, not
all permutations of the Potts labels necessarily preserve
the interaction Jacobian.

\noindent To characterize the symmetries preserved by the oscillator
representation, consider the transformations
\begin{equation}
R_a(s)=s+a\pmod q,
\qquad
M_a(s)=a-s\pmod q,
\end{equation}
where $a\in\{0,\ldots,q-1\}$. These transformations
correspond to rotations and reflections of the equally
spaced phase states on the unit circle. Together, they
form the dihedral group $D_q$, the symmetry group of
a regular $q$-gon.

\noindent Since the interaction coefficient satisfies
\begin{equation}
\kappa_q(r_{ij})=\kappa_q(-r_{ij}),
\end{equation}
all rotations and reflections preserve the interaction
coefficient associated with every edge. Consequently,
\begin{equation}
B_q(P\mathbf{s})=B_q(\mathbf{s}),
\qquad P\in D_q,
\end{equation}
and hence
\begin{equation}
K_s^{\mathrm{stab}}(P\mathbf{s})
=
K_s^{\mathrm{stab}}(\mathbf{s}).
\end{equation}

For $q=3$, the dihedral group $D_3$ coincides with
the full permutation group $S_3$. Thus, all
permutations of the Potts labels preserve the
interaction Jacobian. The same invariance holds
for the binary case, $q=2$.

For $q\geq4$, however, $D_q$ is a proper subgroup
of $S_q$. In fact, the permutations that preserve
the interaction coefficient for every possible pair
of labels are precisely the rotations and reflections
of the regular $q$-gon. This follows because the
coeffient $\kappa_q(r_{ij})$ distinguishes nearest-neighbor
phase separations ($r_{ij}=\pm1$), whose preservation
requires a permutation to be an automorphism of
the cycle graph $C_q$.

Importantly, this symmetry reduction does not imply
that every permutation outside $D_q$ produces a
different stabilization threshold. Depending on the
graph topology, interaction weights, and particular
Potts configuration, such permutations may leave
the interaction Jacobian unchanged or modify it
without changing its largest eigenvalue. The
Rayleigh--Ritz conditions derived above provide
sufficient criteria for determining when a permutation
actually increases or decreases the stabilization
threshold.

Thus, reproducing a discrete objective function does not guarantee that its continuous physical realization preserves the full symmetry of that objective. For the OPM dynamics considered here, the resulting dynamical inequivalence is configuration-dependent, even though the underlying reduction in symmetry is a structural property of the oscillator representation.

\section{Conclusion}
In this work, we investigated the dynamical stability of a specific multi-harmonic oscillator-based Potts machine. We derived a configuration-dependent stability condition from the Jacobian spectrum, which determines the regularization strength required to stabilize a given Potts configuration. We further showed that configurations with the same Potts energy do not necessarily exhibit the same dynamical behavior. In particular, for \(q\geq4\), permutations of the Potts labels can alter the interaction Jacobian and its spectrum, resulting in different stabilization thresholds. While the present work focuses on a specific multiharmonic OPM formulation, these results motivate a broader investigation of the stability properties of other dynamical system formulations of Potts machines~\cite{Potts_model}.

\section*{Acknowledgments}
This material was based upon work supported by the National Science Foundation (NSF) under Grant No. 2328961 and was supported in part by funds from federal agency and industry partners as specified in the Future of Semiconductors (FuSe) program.
\\

\textbf{Author Contributions:} \textbf {E.M.H.E.B Ekanayake:} Conceptualization (equal); Formal analysis (equal); Software (equal); Writing – original draft (equal); Writing – review \& editing (equal). \textbf {Nikhil Shukla:} Conceptualization (equal); Funding acquisition (lead); Supervision (lead); Validation (equal); Writing – review \& editing (equal). \\

\section*{Data availability}
The data that support the findings of this study are available
from the corresponding author upon reasonable request.

\renewcommand{\appendixname}{\MakeUppercase{Appendix}}
\appendix

\section{Stochastic OPM Dynamics and Numerical Simulation}
\label{app:stochastic_opm}

The stability analysis presented in the main text is derived from the
deterministic OPM dynamics in Eq.~\eqref{eq:opm_deterministic}. In the
numerical simulations, however, we also introduce noise to promote exploration of the phase space and reduce trapping in metastable configurations. The resulting stochastic OPM
dynamics are written as

\begin{equation}
\begin{aligned}
\mathrm{d}\theta_i
={}&
\Bigg[
-K\sum_{j=1}^{N} J_{ij}
\sum_{m=1}^{q-1}
m(q-m)
\sin\!\left[m(\theta_i-\theta_j)\right]
\\
&\qquad
-K_s\sin(q\theta_i)
\Bigg]\mathrm{d}t
+\mathrm{d}B_i(t).
\end{aligned}
\label{eq:stochastic_opm}
\end{equation}
where $\mathrm{d}B$ is Brownian noise.

\section{Derivation of the interaction coefficient}
\label{app:kappa_derivation}

In this section, we derive the closed-form expression for the
interaction coefficient 
\begin{equation}
\kappa_q(r_{ij})
=
\sum_{m=1}^{q-1}
m^2(q-m)
\cos\!\left(\frac{2\pi m r_{ij}}{q}\right),
\qquad
r_{ij}\in\mathbb{Z}_q.
\label{eq:app_kappa_def}
\end{equation}
specified in Eq.~\eqref{eq:kappa_closed_form}. We consider separately the cases \(r_{ij}=0\) and \(r_{ij}\neq0\).

\subsection{Case \(r_{ij}=0\)}

\noindent For \(r_{ij}=0\), the cosine factor is unity, and therefore
\begin{align}
\kappa_q(0)
&=
\sum_{m=1}^{q-1}m^2(q-m)
\nonumber\\
&=
q\sum_{m=1}^{q-1}m^2
-
\sum_{m=1}^{q-1}m^3.
\end{align}
Using the standard finite-sum identities
\begin{equation}
\sum_{m=1}^{q-1}m^2
=
\frac{q(q-1)(2q-1)}{6},
\qquad
\sum_{m=1}^{q-1}m^3
=
\frac{q^2(q-1)^2}{4},
\end{equation}
we obtain
\begin{align}
\kappa_q(0)
&=
\frac{q^2(q-1)(2q-1)}{6}
-
\frac{q^2(q-1)^2}{4}
\nonumber\\
&=
\frac{q^2(q^2-1)}{12}.
\label{eq:app_kappa_zero}
\end{align}

\subsection{Case \(r_{ij}\neq0\)}

\noindent For \(r_{ij}\neq0\), define the \(q^{\text{th}}\) root of unity
\begin{equation}
z=e^{2\pi i r_{ij}/q},
\end{equation}
so that \(z^q=1\) and \(z\neq1\). 

Since
\(\cos(2\pi mr_{ij}/q)=\operatorname{Re}(z^m)\), Eq.~\eqref{eq:app_kappa_def}
can be written as
\begin{equation}
\kappa_q(r_{ij})
=
\operatorname{Re}
\left[
q\sum_{m=1}^{q-1}m^2z^m
-
\sum_{m=1}^{q-1}m^3z^m
\right].
\label{eq:app_complex_kappa}
\end{equation}
To evaluate these sums, introduce
\begin{equation}
F(z)
=
\sum_{m=0}^{q-1}z^m
=
\frac{1-z^q}{1-z},
\end{equation}
together with the operator $
D=z\frac{\mathrm{d}}{\mathrm{d}z}.$
Repeated application of \(D\) gives
\begin{equation}
D^nF(z)
=
\sum_{m=0}^{q-1}m^n z^m.
\end{equation}
Evaluating the second and third derivatives at a nontrivial
\(q^{\text{th}}\) root of unity, \(z^q=1\) with \(z\neq1\), gives
\begin{align}
D^2F
&=
\frac{q^2}{z-1}
-
\frac{2qz}{(z-1)^2},
\\
D^3F
&=
\frac{q^3}{z-1}
-
\frac{3q^2z}{(z-1)^2}
+
\frac{3qz(z+1)}{(z-1)^3}.
\end{align}
Hence,
\begin{align}
qD^2F-D^3F
&=
\frac{q^2z}{(z-1)^2}
-
\frac{3qz(z+1)}{(z-1)^3}.
\label{eq:app_difference}
\end{align}
Let
\begin{equation}
\phi=\frac{2\pi r_{ij}}{q},
\qquad z=e^{i\phi}.
\end{equation}
Using
\begin{equation}
z-1
=
2i e^{i\phi/2}\sin\!\left(\frac{\phi}{2}\right),
\end{equation}
the first term in Eq.~\eqref{eq:app_difference} becomes
\begin{equation}
\frac{z}{(z-1)^2}
=
-\frac{1}
{4\sin^2(\phi/2)},
\end{equation}
which is purely real. Similarly,
\begin{equation}
\frac{z(z+1)}{(z-1)^3}
=
\frac{i\cos(\phi/2)}
{4\sin^3(\phi/2)},
\end{equation}
which is purely imaginary and therefore does not contribute to
\(\kappa_q(r_{ij})\). Taking the real part of
Eq.~\eqref{eq:app_difference} therefore yields
\begin{equation}
\kappa_q(r_{ij})
=
-\frac{q^2}
{4\sin^2(\pi r_{ij}/q)},
\qquad r_{ij}\neq0.
\label{eq:app_kappa_nonzero}
\end{equation}
Combining Eqs.~\eqref{eq:app_kappa_zero} and
\eqref{eq:app_kappa_nonzero}, we obtain
\begin{equation}
\kappa_q(r_{ij})
=
\begin{cases}
\dfrac{q^2(q^2-1)}{12},
& r_{ij}=0,\\[8pt]
-\dfrac{q^2}
{4\sin^2(\pi r_{ij}/q)},
& r_{ij}\neq0.
\end{cases}
\end{equation}

\section{Numerical Evaluation of the Rayleigh--Ritz Condition for the
$q=4$ Example}
\label{app:q4_rayleigh_ritz}

We provide here the complete evaluation of the Rayleigh--Ritz condition
for the two permutation-equivalent $q=4$ Potts configurations shown in
Figs.~\ref{fig:phase_label}(b) and \ref{fig:phase_label}(c),
\begin{equation}
\bm{s}_1=[0,2,2,1,3,3],
\qquad
\bm{s}_2=[0,1,1,2,3,3].
\end{equation}
The corresponding interaction Jacobians are
\begin{equation}
B_1=
\begin{bmatrix}
-32&4&4&8&8&8\\
4&-8&-20&8&8&8\\
4&-20&-8&8&8&8\\
8&8&8&-32&4&4\\
8&8&8&4&-8&-20\\
8&8&8&4&-20&-8
\end{bmatrix},
\label{eq:app_B1}
\end{equation}
and
\begin{equation}
B_2=
\begin{bmatrix}
-36&8&8&4&8&8\\
8&-4&-20&8&4&4\\
8&-20&-4&8&4&4\\
4&8&8&-36&8&8\\
8&4&4&8&-4&-20\\
8&4&4&8&-20&-4
\end{bmatrix}.
\label{eq:app_B2}
\end{equation}
Their eigenvalue spectra are
\begin{align}
\operatorname{spec}(B_1)
&=
\{-48,-36,-36,0,12,12\},
\\
\operatorname{spec}(B_2)
&=
\{-48,-40,-32,0,16,16\},
\end{align}
such that
\begin{equation}
\lambda_{\max}(B_1)=12,
\qquad
\lambda_{\max}(B_2)=16.
\label{eq:app_lambda_max}
\end{equation}

The change in the interaction Jacobian under the label permutation is
\begin{equation}
\Delta B_q
=
B_2-B_1
=
\begin{bmatrix}
-4&4&4&-4&0&0\\
4&4&0&0&-4&-4\\
4&0&4&0&-4&-4\\
-4&0&0&-4&4&4\\
0&-4&-4&4&4&0\\
0&-4&-4&4&0&4
\end{bmatrix}.
\label{eq:app_delta_B}
\end{equation}

Since the largest eigenvalue $\lambda_{\max}(B_1)=12$ has multiplicity
two, there are two linearly independent dominant eigenvectors. We select
one normalized dominant eigenvector for the following calculation,
\begin{equation}
\begin{aligned}
\bm v_1
&=
\begin{bmatrix}
0 &
-0.6995 &
0.6995 &
0 &
-0.1034 &
0.1034
\end{bmatrix}^{T},\\
B_1\bm v_1
&=
12\bm v_1.
\end{aligned}
\label{eq:app_v1}
\end{equation}
Applying the Jacobian perturbation to this eigenvector gives
\begin{align}
\Delta B_q\bm v_1
&=
\begin{bmatrix}
-4&4&4&-4&0&0\\
4&4&0&0&-4&-4\\
4&0&4&0&-4&-4\\
-4&0&0&-4&4&4\\
0&-4&-4&4&4&0\\
0&-4&-4&4&0&4
\end{bmatrix}
\begin{bmatrix}
0\\
-0.6995\\
0.6995\\
0\\
-0.1034\\
0.1034
\end{bmatrix}
\nonumber\\
&\approx
\begin{bmatrix}
0\\
-2.7980\\
2.7980\\
0\\
-0.4136\\
0.4136
\end{bmatrix}
=
4\bm v_1.
\label{eq:app_delta_B_v1}
\end{align}
Therefore,
\begin{align}
\bm v_1^{T}\Delta B_q\bm v_1
&=
\bm v_1^{T}(4\bm v_1)
\nonumber\\
&=
4\|\bm v_1\|^2
\nonumber\\
&=
4>0.
\label{eq:app_rayleigh_v1}
\end{align}
The sufficient condition for an increase in the dominant eigenvalue is
therefore satisfied. Applying the Rayleigh--Ritz inequality gives
\begin{align}
\lambda_{\max}(B_2)
&\geq
\lambda_{\max}(B_1)
+
\bm v_1^{T}\Delta B_q\bm v_1
\nonumber\\
&=
12+4
\nonumber\\
&=
16.
\label{eq:app_rayleigh_bound}
\end{align}
Since direct evaluation gives $\lambda_{\max}(B_2)=16$, the
Rayleigh--Ritz bound is attained exactly:
\begin{equation}
\lambda_{\max}(B_2)
=
\lambda_{\max}(B_1)
+
\bm v_1^{T}\Delta B_q\bm v_1
=
12+4
=
16.
\label{eq:app_rayleigh_equality}
\end{equation}

Thus, in this example, the Potts-label permutation produces a positive
Rayleigh quotient,
\begin{equation}
\bm v_1^{T}\Delta B_q\bm v_1=4>0,
\end{equation}
which increases the dominant interaction-Jacobian eigenvalue from
$12$ to $16$.

\def\bibsection{\section*{References}}
\bibliography{References}

\end{document}